\documentclass[12pt]{article}
\usepackage[dvips]{color}
\usepackage{mathtext}
\usepackage[koi8-r]{inputenc}
\usepackage{graphicx}
\usepackage[figuresright]{rotating}
\usepackage{amssymb,eucal}
\usepackage{cite}
\date{}

\newcommand{\beq}{\begin{equation}}
\newcommand{\eeq}{\end{equation}}
\newcommand{\beqn}{\begin{eqnarray}}
\newcommand{\eeqn}{\end{eqnarray}}  

\begin{document}
\title{Test of exotic scalar and tensor interactions in $K_{e3}$
decay using stopped positive kaons}

\author{
A.S.~Levchenko$^{a}$, A.N.~Khotjantsev$^{a}$, A.P.~Ivashkin$^{a}$,
M.~Abe$^{b}$, \\
M.A.~Aliev$^{a}$, V.V.~Anisimovsky$^{a}$, M.~Aoki$^{c}$, 
I.~Arai$^{b}$, \\
Y.~Asano$^{b}$, T.~Baker$^{c,d}$, M.~Blecher$^{e}$, 
P.~Depommier$^{f}$,\\
M.~Hasinoff$^{g}$, K.~Horie$^{h}$, H.C.~Huang$^{i}$,
Y.~Igarashi$^{c}$,\\ T.~Ikeda$^{j}$, J.~Imazato$^{c}$, M.M.~Khabibullin$^{a}$,
 Y.G.~Kudenko$^{a}$,\\ 
Y.~Kuno$^{c,h}$, L.S.~Lee$^{k}$, G.Y.~Lim$^{k,c}$, J.A.~Macdonald$^{l}$, \\
 D.R.~Marlow$^{m}$, C.R.~Mindas$^{m}$,
O.V.~Mineev$^{a}$, C.~Rangacharyulu$^{d}$, \\ S.~Shimizu$^{h}$, 
Y.-M.~Shin$^{d}$, A.~Suzuki$^{b}$, A.~Watanabe$^{b}$,\\ 
N.V.~Yershov$^{a}$, T.~Yokoi$^{c}$\\}
 \maketitle 
 \begin{center}
(E246 KEK--PS COLLABORATION) 
\end{center}
\begin{center}
 $^{a}$Institute for Nuclear Research of RAS, Moscow 117312, Russia;\\ 
 $^{b}$University of Tsukuba, 305-0006, Japan;\\ 
 $^{c}$KEK, 305-0801, Japan;\\
 $^{d}$University of Saskatchewan, S7N 0W0, Canada;\\
 $^{e}$Virginia Polytechnic Institute and State University, USA;\\
 $^{f}$University of Montreal, H3C 3J7, Canada;\\ 
 $^{g}$University of British Columbia, Vancouver, V6T 1Z1 Canada;\\
 $^{h}$Osaka University, 560-0043, Japan;\\
 $^{i}$National Taiwan University, Taipei, Taiwan;\\
 $^{j}$Institute of Physical and Chemical Research, Hirosawa 2-1, Japan;\\
 $^{k}$Korea University, Seoul 136-701, Korea;\\
 $^{l}$TRIUMF, V6T 2A3, Canada;\\
 $^{m}$Princeton University, NJ 08544, USA.\\
 \end{center}

{\bf Abstract}\\
The form factors of the decay $K^+ \to \pi^0e^+\nu$ ($K_{e3}$) 
have been determined from the comparison of  
the experimental and Monte Carlo Dalitz distributions containing 
 about  $10^5$ $K_{e3}$ 
events.  The following values of the parameters were obtained: 
$\lambda_+ = 0.0278 \pm 0.0017(stat) \pm 0.0015(syst)$, 
$f_S/f_+(0) = 0.0040 \pm 0.0160(stat) \pm 0.0067(syst)$ and 
$f_T/f_+(0) = 0.019 \pm 0.080(stat) \pm 0.038(syst)$. Both scalar $f_S$ 
and 
tensor $f_T$ form factors  are
consistent with the Standard Model predictions of zero values. 

\section{Introduction}

The most general Lorentz invariant  form of 
the matrix element 
of  $K_{e3}$ decay can be written as~\cite{steiner,group}:
\begin{eqnarray}
M & \propto & f_{+}(q^2)({P}_{K} + {P}_{\pi^0})^{\lambda}
\bar{u}_{e}\gamma_{\lambda}(1-\gamma_5)u_{\nu} \nonumber \\
&& + f_{-}(q^2) m_l\bar{u}_{e}(1-\gamma_5)u_{\nu}
+ 2m_Kf_S\bar{u}_{e}(1-\gamma_5)u_{\nu} \nonumber \\
&& + (2f_T/m_K)(P_K)^\lambda(P_{\pi^0})^\mu\bar{u}_{e}
\sigma_{\lambda\mu}(1-\gamma_5)u_\nu,  
\label{eq:main}
\end{eqnarray}
where $f_{\pm}(q^2)$, $f_S$ and $f_T$ are  the vector, scalar and tensor 
 form-factors, $P_K$ and $P_{\pi}$ are four-momenta of
the $K^+$ and $\pi^0$.
The vector form-factors $f_{\pm}(q^2)$ are 
assumed to be linearly dependent on the momentum transfer squared
$q^2 = (P_K-P_{\pi})^2$ and they can be represented by the equation $f_{\pm}(q^2) =
f_{\pm}(0)(1\pm\lambda_{\pm}(q^2/m_{\pi^0}^2))$.  Because of the small mass of the positron
the part of the matrix element that depends on  $f_{-}(q^2)$ is negligible and
there are just three free parameters of the theory: $\lambda_+$, $f_S$ and 
$f_T$. Within the Standard Model (SM), due to 
$W$--boson exchange,
no terms  
other than those of a pure vector nature are expected.  A possible
contribution to  $f_S$ and $f_T$ from electroweak radiative corrections
is negligibly small and, therefore,  nonzero values of  $f_S$ and $f_T$ 
would signal
new physics beyond the SM. Actually, such deviations from
zero for these form factors were measured in~\cite{steiner,akimenko}, but
were not confirmed in a recent experiment~\cite{ke3}, where $f_S$ and $f_T$
were measured with
sensitivity similar to Refs.~\cite{steiner,akimenko}. 

In this paper, we present results  of a reanalysis of $K_{e3}$ data
taken with the  E246 detector at the KEK  12--GeV  proton synchrotron.
The first result obtained using this data set was published in~\cite{ke3}.  

\section{Experiment}
\begin{figure}[htb]
\centering\includegraphics[width=15cm, angle=0]{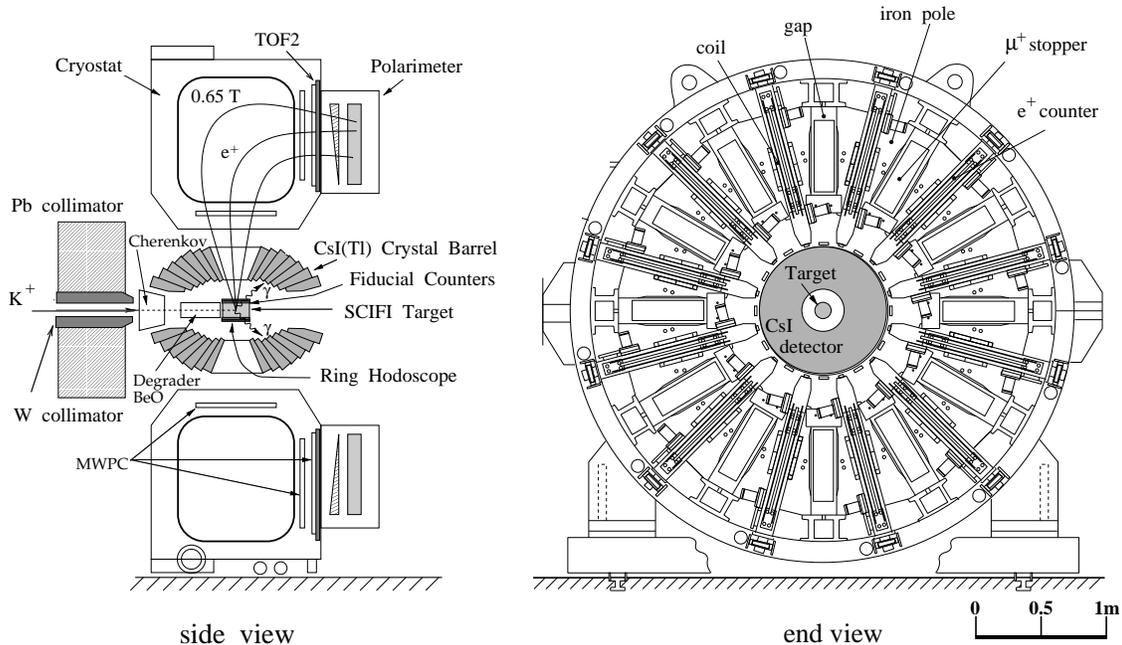} 
\caption{The layout of the E246 set--up}
\label{fig:setup}
\end{figure}

The experiment was performed using the set--up constructed 
to search for T--violation in $K^+\to \pi^0 \mu^+ \nu$ decay. The 
experimental arrangement is shown in Fig.~\ref{fig:setup}, and is
described in detail elsewhere~\cite{abe,csi,iva,cher}. 
Kaons  with $P_{K^+}=660~$MeV/$c$ are identified by a Cherenkov counter,
slowed  in a BeO degrader and then stopped in a target array of 256
scintillating fibers located at the center of a 12-sector  
superconducting toroidal  spectrometer.
Charged particles from  kaon decays in the target were tracked by 
means
of multiwire proportional chambers at the 
entrance (C2) and exit (C3 and C4)  of
each magnet sector, along with the target 
and a scintillation ring hodoscope around the target. The momentum 
resolution of $\sigma_{p}=2.6~$MeV/$c$ at $p=205~$MeV/$c$ was obtained using 
mono-energetic products from the two-body decay $K^+\to \pi^+\pi^0$.

The energies and angles
of  the photons from $\pi^0$  decays are measured  by a CsI(Tl) photon
detector consisting of 768 modules~\cite{csi}. The photon detector covers a 
solid angle of $3\pi$ steradians, with openings for the beam entry and 
exit and 12 holes for charged particles to pass into the magnet 
gaps. The
photon energy  was obtained by summing  the cluster energy distributed among
several crystals surrounding a central  crystal. The position of
the photon cluster  is determined 
using 
an energy-weighted ``center-of-gravity" method. To suppress  accidental background from 
the beam, 
timing information from each crystal was used. A time resolution 
of 3.8 nsec ($\sigma$) for the wide photon energy range of 10--220 MeV
was achieved. The energy 
resolution was obtained to be  $\sigma_{E_{\gamma}}/E_{\gamma} \simeq 3.0\%$ 
at $E_{\gamma}=150~$MeV. The invariant mass resolution of 
$\sigma_{\gamma\gamma} = 7.5 $ MeV/$c^2$ at $M_{\gamma\gamma} = 133.1$
MeV/$c^2$,
and angular resolution
of $\sigma_{\theta}=2.4^{\circ}$
were obtained in the experiment~\cite{csi, marat}. 

The study of the $K_{e3}$  decay required  a trigger condition different from
that of the main experiment. The following trigger was used to accumulate
$K_{e3}$ events
\begin{equation}
C_{K} \times Fid_i \times TOF2_{i} \times 2\gamma \times GapVeto.
\label{eq:trigger}
\end{equation}
The Cherenkov condition $C_K$ ensures that the beam particle is a $K^+$.
The coincidence between $C_K$, delayed by 2 nsec, and the fiducial
counters ($Fid_i$) for the charged particles eliminates the triggers
due to decays in flight. The time of flight signal from the corresponding gap
($TOF2_i$) completes the charged particle trigger condition. 
The trigger also
required two hits in the CsI ($2\gamma$).
The requirement $GapVeto$ eliminates events where the charged particle 
may have lost energy
in the CsI crystals
around the $i^{th}$ hole.
 
The $K_{e3}$ data were collected for two spectrometer field 
settings, B=0.65 and B=0.9 Tesla. 
 
\section {Analysis }

\begin{figure}[htbp]
\begin{center}
{\centering\includegraphics[width=15cm, angle=0]{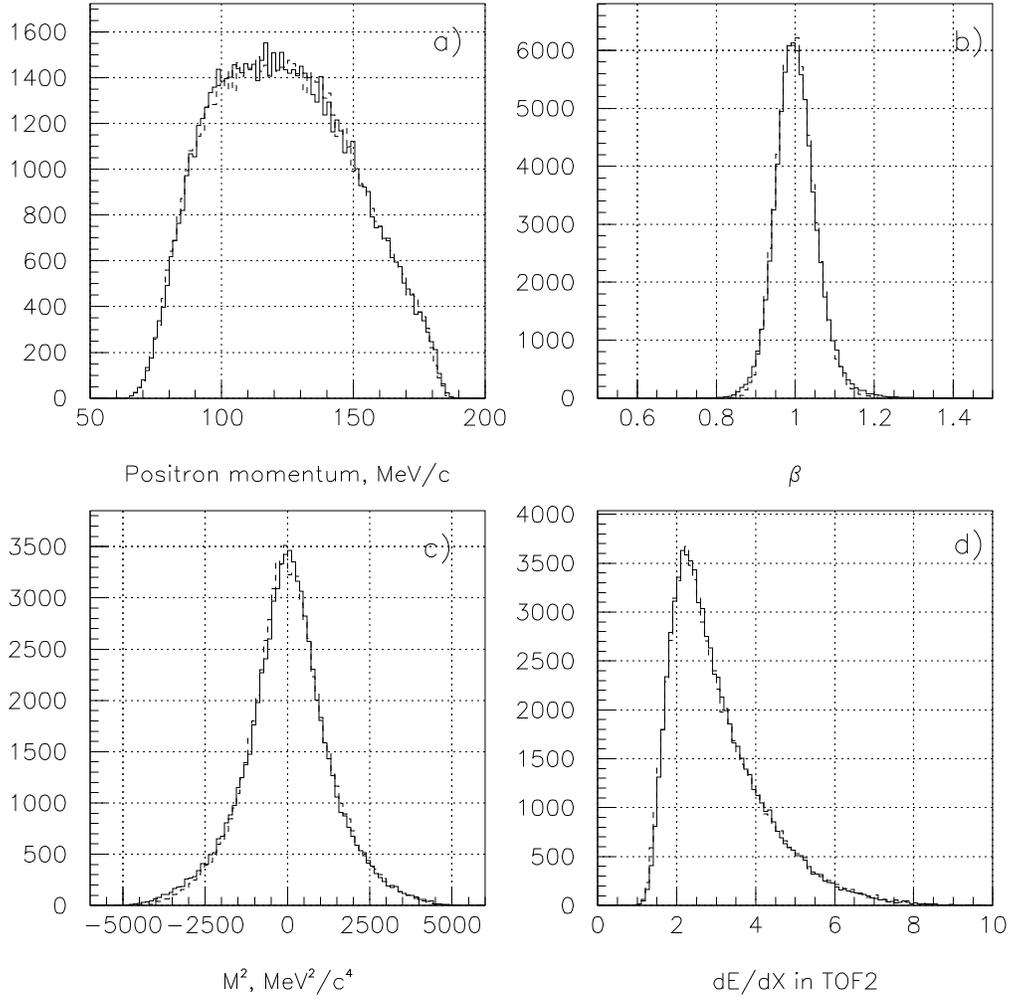}} 
\end{center}
\caption{(a) Momentum spectrum of $e^+$ from $K_{e3}$ decay;
(b) $\beta$ spectra of $e^+$ ($\beta = {L \over \tau \cdot c} $);
(c) reconstructed TOF mass squared of $e^+$;  
(d) energy deposit of $e^+$ in TOF2 counter.
Solid  line corresponds to experimental and dashed line to
GEANT--simulated events.
}
\label{fig:eplusprop}
\end{figure}

$K_{e3}$ events were selected using  the following  requirements.
 A clean hit  pattern in the target and a delayed decay at least 2 nsec after
the $K^+$ arrival time measured by Cherenkov counter suppressed
the $K^+$ decays-in-flight. 
Cuts on 
the charged-particle momentum, $p<190~$MeV/$c$,  and the opening angle 
between $e^+$ and $\pi^0$, $\theta_{e^+\pi^0}<170^{\circ}$, effectively 
removed the  $K_{\pi2}$ events.
$\pi^+$ decays-in-flight were rejected by a cut on the
track reconstruction $\chi^2<14$. The photon conversion 
events were suppressed by requiring 
single hits in the ring and fiducial
counters and
a single track in the target. 
Events with more than two hits in the electromagnetic 
calorimeter were rejected. The cut on the invariant mass was 
$75<M_{\gamma\gamma}<140~$MeV/$c^2$.
The main  
criterion separating $K_{e3}$ positrons from 
$K_{\mu3}$ muons was 
derived from the time-of-flight (TOF) measurement between the
fiducial counters and scintillating counters (TOF2) located at the exit of 
each spectrometer     
gap (see Fig.~\ref{fig:setup}).  
The mass squared of the charged particle was determined as
\begin{equation}
M^2=P^2((L/\tau \cdot c)^2-1),
\label{eq:tof}
\end{equation}
 where $L$ and $\tau$ are the length and time of
flight of the charged particle between fiducial and TOF2 counters,
 and $c$ is the
speed of light in vacuum. The TOF resolution is
$\sigma_{TOF}=300$~ps, and the mass squared spectrum is shown in 
Fig.~\ref{fig:eplusprop}.
To separate positrons from muons, a cut on the
mass squared $-4500<M^2_{TOF}<4500~$MeV$^2$/$c^4$ was chosen.

Compared to the analysis of~\cite{ke3}, improvements in reconstruction of
the charged particle track were made. In the new analysis, information about 
the kaon stopping position from the target $(x,y)$ and  
ring counter $(z)$ was included in the momentum
reconstruction routine. The reconstruction of the $\pi^0$ kinematics
was also improved by using 
tighter time windows in the CsI and 
improved calibration.  The Monte Carlo routines properly included
the correct kaon stopping distribution in the target and the electromagnetic
shower leakage effects in the CsI.  This allowed us to use all CsI 
crystals
for reconstruction of the $\pi^0$, while in Ref.~\cite{ke3} 
events were rejected where either photon hit a crystal adjacent to 
the 12 charged-partilcle holes.
 Overall optimization resulted in approximately a factor of four increase
 in acceptance.
For 0.65-T magnetic field we extracted  102k 
good  $K_{e3}$ events
 with  background  contamination of 
0.21$\%$ from $K_{\mu3}$ and 0.34$\%$  from  $K_{\pi2}$ decays. 
 
\begin{figure}[htbp]
\begin{center}
{\centering\includegraphics[width=15cm, angle=0]{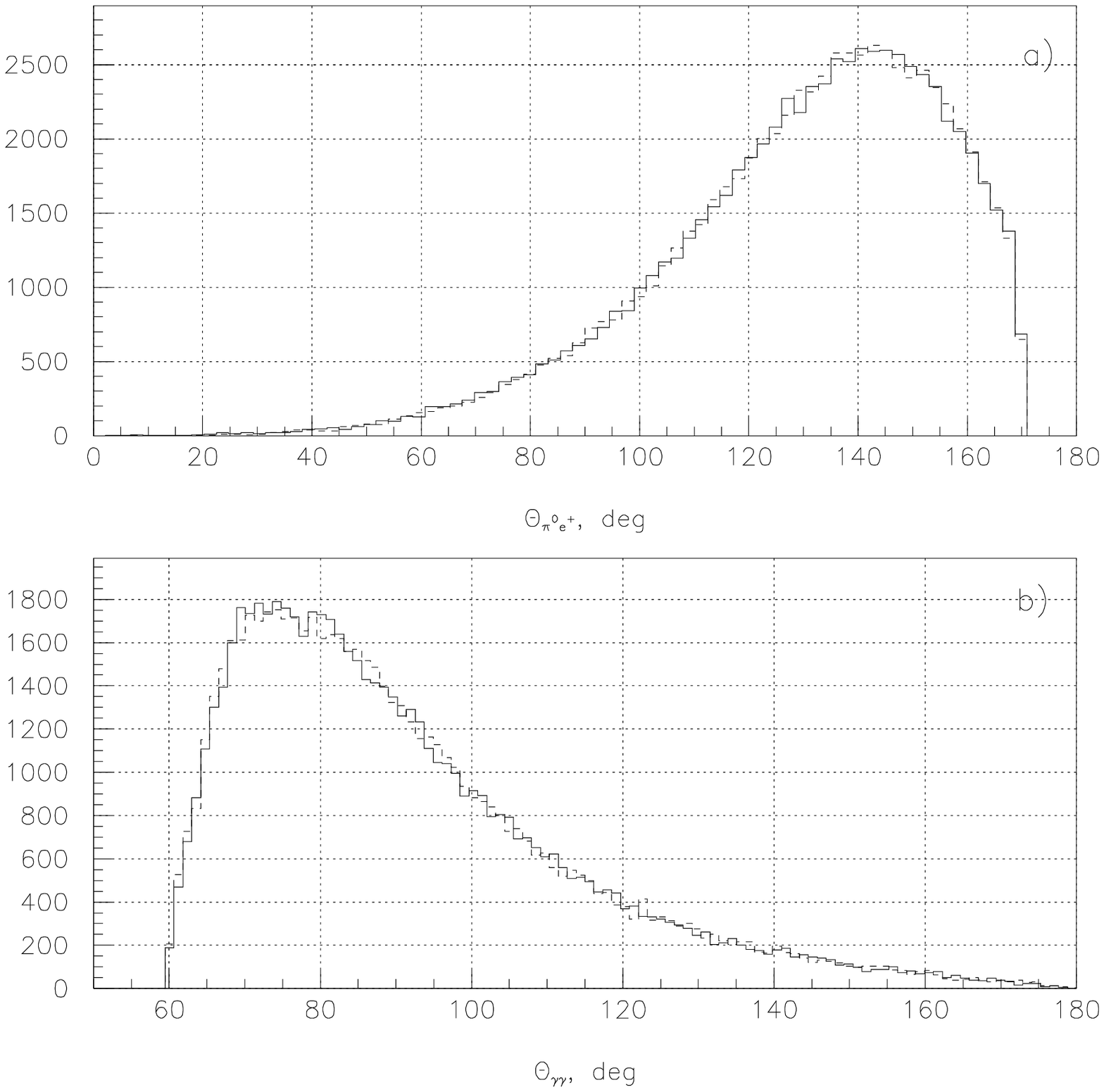}} 
\end{center}
\caption{ $K_{e3}$ decays:
a) angle between $e^+$ and $\pi^0$;
b) angle between photons $\theta_{\gamma\gamma}$. Solid line - experiment, 
dashed line - Monte Carlo.}
\label{fig:angles_ke3}
\end{figure}

\begin{figure}[htbp]
\begin{center}
{\centering\includegraphics[width=15cm,  angle=0]{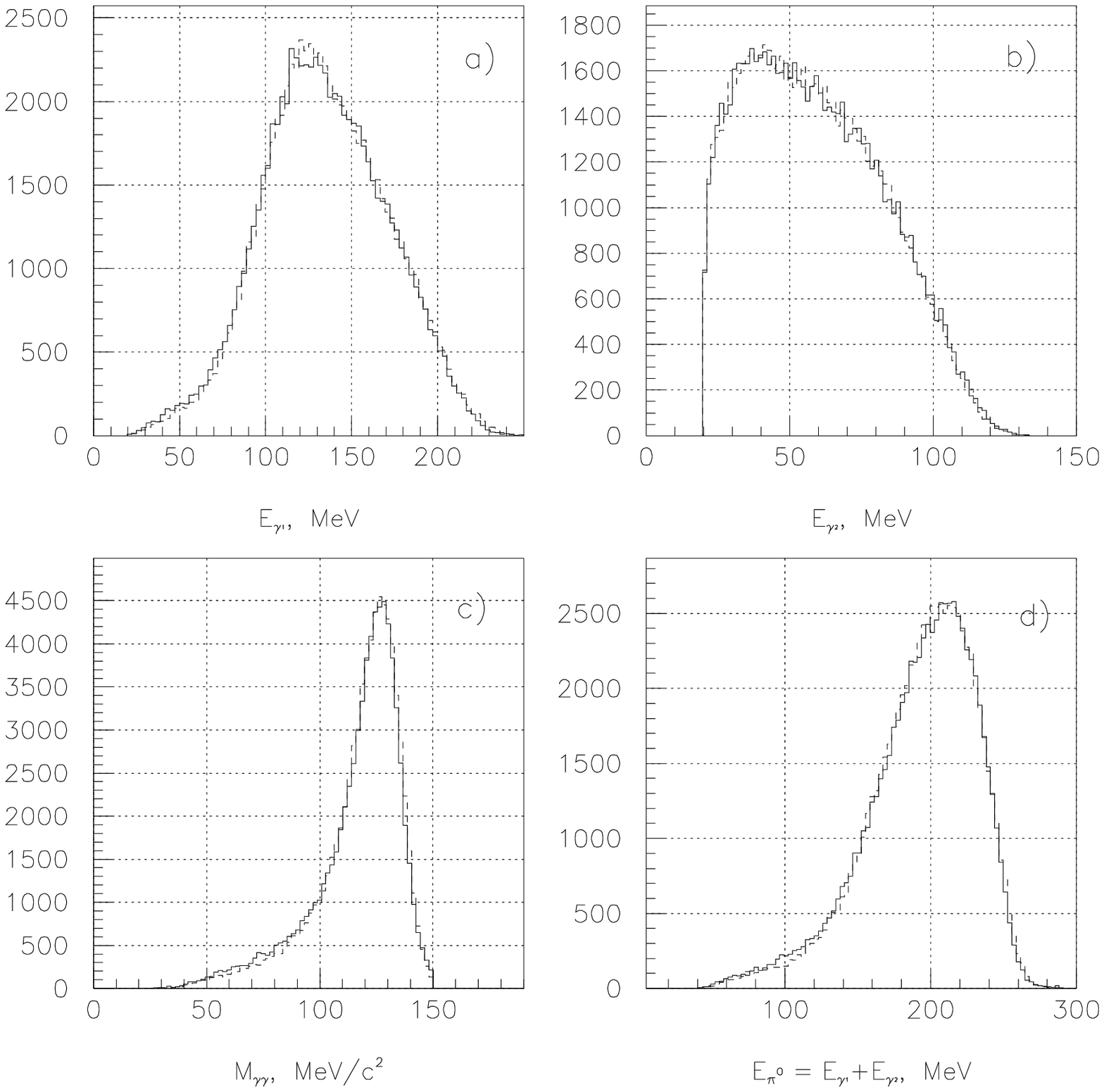}} 
\end{center}
\caption{ $K_{e3}$ decays:
a) energy of $\gamma_1$ ($E_{\gamma_1}>E_{\gamma_2}$);
b) energy of $\gamma_2$;
c) invariant mass $M_{\gamma\gamma}$;
d) energy of $\pi^0$ ($E_{\pi^0}=E_{\gamma_1}+E_{\gamma_2}$).
Solid line - experiment, dashed line - Monte Carlo.}
\label{fig:eg_ke3}
\end{figure}

The distributions of the $e^+ - \pi^0$ opening angle  
and 
the opening angle between photons from $\pi^0\to \gamma\gamma$ for
$K_{e3}$ decay is shown in Fig.~\ref{fig:angles_ke3}.
Energy spectra of the $K_{e3}$ decay are shown in Fig.~\ref{fig:eg_ke3}.

The extraction method for the parameters is based on comparison of the 
 experimental and Monte
Carlo Dalitz distributions. If one 
allows for the existence of the exotic 
interactions, the 
Dalitz plot density is given by~\cite{braun}:
\begin{equation}
\rho(E_e,E_{\pi}) = {f^2_+(q^2)}(A+B{\xi}(q^{2})+C{\xi}^2(q^{2}))
\label{eq:nextmain}
\end{equation}
where 
$$ A = m_K(2E_eE_{\nu}-m_{K}E^{'}_{\pi})+m^2_e({E^{'}_{\pi}/4}-E_{\nu})$$
$$ B=({m^{2}_{e}})(E_{\nu}-{E^{'}_{\pi}/2})$$
$$ C=({m^2_e}){E^{'}_{\pi}/4}$$
$$ E^{'}_{\pi} = ({m^2_K}+m^2_{\pi} - m^2_e)/(2m_K) - E_{\pi} $$
$${\xi}(q^2) = {{(2m_K/m_e)R_S+(m_e/m_K+2(E_{\nu}-E_e)/m_e)R_T} \over 
{(1+(m_e/m_K)R_T)}} $$
\begin{center}
$R_S=f_S/f_+$~~~~~~~~~~~~~$R_T=f_T/f_+$
\end{center}
The  parameters of the decay can be  extracted by selecting the minimum of the
 $\chi^2$ variable defined as
\begin{equation} 
 \chi^2=2 \cdot \sum_{i=1}^n(N^{MC}_i-N^{exp}_i)+
 N_i^{exp}\cdot log\bigl({N^{exp}_i \over N^{MC}_i}\bigr),
\end{equation}
where $n$ is the number of bins over the Dalitz plot, 
$N^{MC}_i$ the number of Monte Carlo 
events in each bin, and
 $N^{exp}_i$ the number of experimental events in each bin.
 
\begin{figure}[htbp]
\begin{center}
{\centering\includegraphics[width=16cm, angle=0]{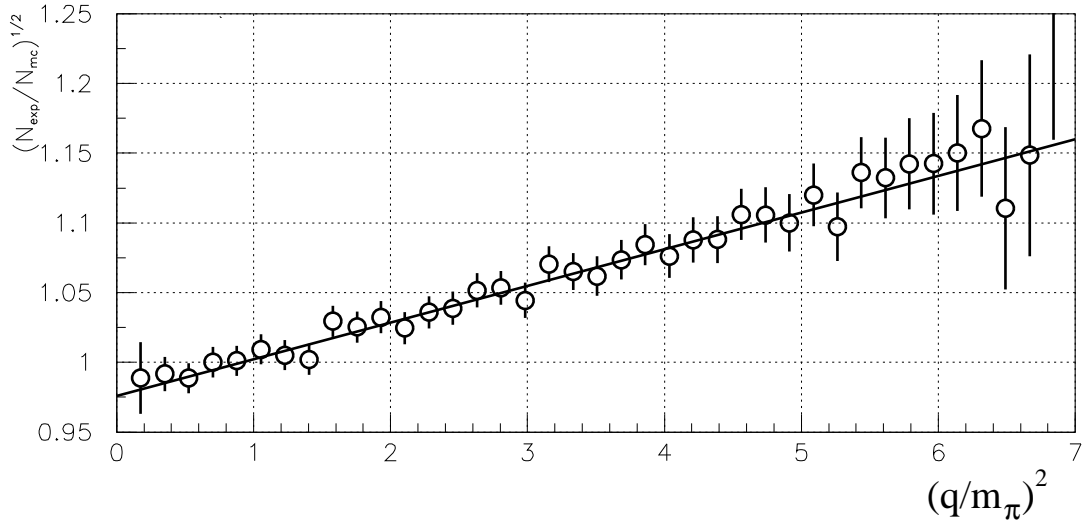}} 
\end{center}
\caption{The  $q^2$ dependence of the ratio $N_{exp}/N_{mc}$.
The straight line shows a linear fit with  $\lambda_+=0.0278$.}
\label{fig:vminusa}
\end{figure}

  Within the Standard Model the only 
parameter to be obtained is $\lambda_+$. It
can be extracted in a model--independent way  
 from the   $q^2$ dependence of $f_+$. Using eq.~\ref{eq:nextmain},
$\sqrt{\rho(E_e,E_{\pi})} \propto f_+=f_+(0)(1+\lambda_+q^2/m_{\pi^0}^2)$,
and, therefore,  
$\lambda_+$ can be extracted from the ratio
\begin{equation}
(N_{exp}/N_{mc}(\lambda_+=0))^{1/2}=(1+\lambda_+q^2/m_{\pi^0}^2).
\end{equation}
 The  $q^2$ dependence of this  ratio
 shown in Fig.~\ref{fig:vminusa} allows us to determine the value 
$\lambda_+=0.0278\pm0.0016(stat)$. 

  For extraction 
of all three parameters $\lambda_+$, $f_S$ and $f_T$, 
the Dalitz distribution fit was performed 
using the $P_{e^+}$, $\theta_{\pi^0e^+}$  
variables. The angle between the pion and positron,
 $\theta_{\pi^0e^+}$,  was preferred instead of pion energy  $E_{\pi^0}$ 
in order  to reduce a systematic
error related to the energy leakage in the electromagnetic calorimeter. 
The bin sizes chosen were $\Delta P_{e^+} = 3.125$~MeV/$c$ and 
$\Delta \theta_{e^+\pi^0} = 4.2^{\circ}$. Radiative corrections to 
the Dalitz plot were taken
into account  according to Ginsberg
\cite{ginsberg}. 
The $\chi^2$ between
experimental and Monte Carlo Dalitz distributions was minimized by a 
program based on
MINUIT~\cite{minuit}. The  values obtained for $\lambda_+$,
and the scalar and tensor form factors  are
\begin{eqnarray}
\lambda_+ = 0.0278 \pm 0.0017(stat) \nonumber \\
f_S/f_+(0) = 0.004 \pm 0.016(stat)   \nonumber \\
f_T/f_+(0) = 0.019 \pm 0.080(stat).  
\end{eqnarray}
Since our analysis relies on the proper detector
responses 
used by the Monte Carlo simulation,  comparisons and checks were made
wherever possible. For example, to check the values obtained
 and to see possible unknown systematic 
effects, a similar  analysis of the scatter 
plot $\theta_{\gamma\gamma}$ vs. $P_{e^+}$ 
was also performed.   From this approach 
it was   found that  
 $\lambda_+ = 0.0281 \pm 0.0017(stat)$,  
 $f_S/f_+(0) = 0.001 \pm 0.018(stat)$, and  
 $f_T/f_+(0) = 0.007 \pm 0.100(stat) $. No significant variations of 
 the  $\lambda_+$ value  were seen, and values of the form factors were 
 consistent with zero within a 1 $\sigma$ uncertainty. 

The main sources of systematic errors 
were related to
detector inefficiencies, misalignments  of the detector elements,  background 
contamination, and uncertainties connected to the Monte Carlo simulation. 
  All
these sources were studied and the estimations of the   
systematic errors are
presented in  Table~\ref{tab:syst}. 
\begin{table}[htb]
\caption{Systematic errors.}
\begin{center}
\begin{tabular}{l c c c }
\hline
&&& \\
Source & $\lambda_+$ & $f_S$ & $f_T$ \\
&&& \\
\hline
&&& \\
MWPC's misalignment~~~~~~~~~~~~~~~           & 0.0006& 0.0010& 0.0022    \\
MWPC's spatial resolution     & 0.0002  & 0.0010& 0.0010    \\
$e^+$ identification          & 0.0007  & 0.0022& 0.0004   \\
Bremsstrahlung                & 0.0010  & 0.0015& 0.0280   \\
$\chi^2$ cut point            & 0.0001  & 0.0010& 0.0018    \\
CsI barrel misalignment       & 0.0005  & 0.0042& 0.0240    \\
$E_{\gamma}$ and $M_{\gamma\gamma}$ cut point
                              & 0.0003  & 0.0020& 0.0027    \\
$\gamma \rightarrow e^+e^-$ conversion        
                              & 0.0001  & 0.0018& 0.0100      \\
Pile-up in the CsI            & 0.0004  & 0.0013& 0.0015    \\
&&& \\
Total                         & 0.0015  & 0.0067& 0.0380     \\
&&& \\
\hline
\end{tabular}
\end{center}
\label{tab:syst}
\end{table}
\section{Result}
The result presented here is based on about $10^5$ good $K_{e3}$ 
events. We have obtained 
\begin{eqnarray}
\lambda_+ = 0.0278 \pm 0.0017(stat) \pm 0.0015(syst) \nonumber \\
f_S = 0.0040 \pm 0.0160(stat) \pm 0.0067(syst)  \nonumber \\
f_T = 0.019 \pm 0.080(stat) \pm 0.038(syst). 
\end{eqnarray}
Using only
one data set at B=0.65 T, we have improved statistical errors by a factor 
of 1.5  and systematic
errors  for
$\lambda_+$, $f_S$ and $f_T$ were reduced by the factors 2.0, 2.1, and 2.4,
respectively, compared to our previous result~\cite{ke3}. 
This result is in agreement with the Standard Model prediction and
there is no evidence for  a deviation
from zero    for the values of scalar and tensor form factors. 
We expect  to further improve our accuracy after completing the 
analysis of
the $K_{e3}$ data accumulated at B=0.9~T.

\end{document}